\documentclass[letterpaper] {article}     
\usepackage{cite}               
\usepackage{epsfig}

\newcommand{\lant}{{\stackrel{\rightarrow}{\scriptscriptstyle \leftarrow}}}
\newcommand{\lpar}{{\stackrel{\rightarrow}{\scriptscriptstyle \rightarrow}}}

\begin{document}
\bibliographystyle{unsrt}

\title{
{The $Q^2$-dependence of the Generalised Gerasimov-Drell-Hearn \\
Integral for the Proton}}
\date{August 17, 2000}

\maketitle              
\begin{center}          

{
{ HERMES Collaboration}\medskip\\ 
A.~Airapetian$^{\mbox{ ae}}$,
N.~Akopov$^{\mbox{ ae}}$,
I.~Akushevich$^{\mbox{ g}}$,
M.~Amarian$^{\mbox{ w,z,ae}}$,
J.~Arrington$^{\mbox{ b}}$,
E.C.~Aschenauer$^{\mbox{ g,m,w}}$,
H.~Avakian$^{\mbox{ k}}$,
R.~Avakian$^{\mbox{ ae}}$,
A.~Avetissian$^{\mbox{ ae}}$,
E.~Avetissian$^{\mbox{ ae}}$
% ...
} \\
{\it        
% ...
$^{\mbox{ b}}$Physics Division, Argonne National Laboratory, Argonne, Illinois 60439-4843, USA\\
% ...
$^{\mbox{ g}}$DESY Zeuthen, 15738 Zeuthen, Germany\\
% ...
$^{\mbox{ k}}$Istituto Nazionale di Fisica Nucleare, Laboratori Nazionali di Frascati, 00044 Frascati, Italy\\
% ...
$^{\mbox{ m}}$Department of Subatomic and Radiation Physics, University of Gent, 9000 Gent, Belgium\\
% ...
$^{\mbox{ w}}$Nationaal Instituut voor Kernfysica en Hoge-Energiefysica (NIKHEF), 1009 DB Amsterdam, The Netherlands\\
% ...
$^{\mbox{ z}}$Istituto Nazionale di Fisica Nucleare, Sezione Sanit\`a and Physics Laboratory, Istituto Superiore di Sanit\`a, 00161 Roma, Italy\\
% ...
$^{\mbox{ ae}}$Yerevan Physics Institute, 375036, Yerevan, Armenia\\
} 

\end{center}           

\begin{abstract} 
The dependence on $Q^2$ (the negative square of the 4-momentum of
the exchanged virtual photon) of the generalised 
Gerasimov-Drell-Hearn integral for the proton has been measured in the range 
1.2~GeV$^2<Q^2<12$~GeV$^2$
by scattering longitudinally polarised positrons
on a longitudinally polarised hydrogen gas target. 
The contributions of the nuc\-leon-res\-on\-ance
and deep-inelastic regions 
to this integral have been evaluated separately.
The latter has been found to dominate for $Q^2 >3$~GeV$^2$, while both contributions are important at low $Q^2$.
The total integral shows no significant deviation from a 1/$Q^2$ behaviour
in the measured $Q^2$ range, and thus no sign of
large effects due to either nuc\-leon-res\-on\-ance excitations or 
non-leading twist. 

\end{abstract}

\twocolumn

The Gerasimov-Drell-Hearn (GDH) sum rule~\cite{ger} 
relates the anomalous contribution $\kappa$ in the nucleon magnetic moment to 
an energy-weighted integral of the difference of the
nucleon's total spin-dependent photoabsorption cross sections:

\begin{equation}
\int_{\nu_0}^{\infty} \lbrack \sigma_{\frac{1}{2}}( \nu )-\sigma_{\frac{3}{2}}( \nu ) \rbrack
\frac{
d\nu }{\nu}=
-\frac{2\pi^2 \alpha}{M^2} \kappa^2,
\label{gdh}
\end{equation}
where $\sigma_{\frac{1}{2}(\frac{3}{2})}$ is the photoabsorption cross section for 
total helicity of the photon-nucleon system equal to $\frac{1}{2}$~($\frac{3}{2}$), 
$\nu$ is the photon energy in the target rest frame, $\nu_0$ is the pion production
threshold and $M$ is the nucleon mass.
For the proton ($\kappa_{\mathrm p}=+1.79$) the GDH sum rule prediction is 
$-204\,\mu$b.
The importance of this sum rule is due to the fact 
that it is based mostly on very general 
principles of causality, unitarity, crossing symmetry, and Lorentz and gauge invariance.
It has never been directly
tested, due to the need for a circularly polarised beam with a 
longitudinally polarised target, and a wide range of photon energies 
that has to be covered. 
There are available
several predictions for the contribution of nucleon resonance excitation 
to the GDH integral~\cite{RP567}, 
derived from multipole analyses of data for unpolarised single-pion photoproduction, 
and a prediction for the contribution 
of high energy multi-hadron production~\cite{nikolo},
based on a multiple-Reggeon exchange analysis of deep-inelastic asymmetries.
The estimate from multipole analysis was confirmed by
preliminary results from the GDH experiment at Mainz~\cite{MAMI}, which covered 
the photon energy range from 200~MeV  up to 800~MeV.

The GDH integral can be generalised to the case of absorption of polarised
transverse  virtual photons with squared four-momentum $-Q^2$~\cite{RP}:

\begin{eqnarray}
I_{GDH}(Q^2) \equiv
\int_{\nu_{0}}^{\infty} \lbrack \sigma_{\frac{1}{2}}( \nu, Q^2 )-\sigma_{\frac{3}{2}}( \nu, Q^2 ) 
\rbrack
\frac{
d\nu }{\nu}
\label{for2}
\\ 
= 16 \pi^2 \alpha \int_0^{x_{0}}\frac{g_1(x,Q^2)-\gamma^2 g_2(x,Q^2)}
{Q^2 \sqrt{1+\gamma^2}}
dx
\label{for2a}
\\ 
= \frac{8 \pi^2 \alpha}{M}\int_0^{x_{0}}\frac{A_1(x,Q^2) F_1(x,Q^2)}
{K}
\frac{dx}{x},
\label{for2b}
\end{eqnarray}
where $g_1$ and $g_2$ are the polarised structure functions of the nucleon,
$\gamma^2=Q^2/\nu^2=(2Mx)^2/Q^2$, 
$x=Q^2/2M\nu$ and 
$x_{0}=Q^2/2M\nu_{0}$. The quantity $A_1$ is the longitudinal
asymmetry for virtual photoabsorption, while $F_1$ is the unpolarised 
structure function of the nucleon.
The Gilman notation~\cite{gil} for the
virtual photon flux factor $K = \nu \sqrt{1+\gamma^2}$ has been used. 
It should be noted that elastic scattering occurring at $x=1$ does not
contribute to the generalised integral.
Other generalisations of the GDH integral 
also have been considered~\cite{RP}. 
They differ from the definition given in Eq.~\ref{for2} 
by terms in the integral that are proportional to $\gamma^2$ 
and which therefore vanish in both the real-photon ($Q^2$ = 0) and the deep-inelastic 
($Q^2 \gg 1$~GeV$^2$ and $\gamma^2 \rightarrow 0$) limits.
Since $\gamma^2$ is not small in the nuc\-leon-res\-on\-ance region and at moderate
$Q^2$ (e.g. $\gamma^2$ is larger than unity for the $P_{33}(1232)$-resonance
for 0.2~GeV$^2<Q^2 <2$~GeV$^2$), these generalisations are equivalent
for finite values of $Q^2$ only if the contributions of the 
nuc\-leon-res\-on\-ance excitations are small.    

The generalisation of the GDH integral to non-zero photon virtuality $Q^2$ 
provides a way to study the transition 
from polarised lepton scattering from
the nucleon, which is dominated by deep-inelastic scattering (DIS)
at large photon-nucleon centre of mass energy  
$W=\sqrt{M^2+2M\nu-Q^2}$,
to the polarised real photon absorption on the nucleon, which is dominated 
by nuc\-leon-res\-on\-ance excitation at low $W$. 
In leading twist (e.g. for $Q^2\rightarrow \infty$), Eq.~\ref{for2a} simplifies:
the elastic contribution excluded from the integral  is of
higher twist, the factor $1/\sqrt{1+\gamma^2}$ is 1+$\mathcal{O}(1/Q^2)$, and in leading
twist approximation it is generally believed (though not formally proven)
that the Burkhardt-Cottingham sum rule $\int_0^1 g_2(x,Q^2) dx =0$
holds.  This leads to:

\begin{equation}
I_{GDH}(Q^2)_{\gamma^2 \rightarrow 0} = \frac{16\pi^2\alpha}{Q^2}\Gamma_1,
\label{I1}
\end{equation}
where $\Gamma_1 \equiv \int_0^1 g_1(x)dx$ is predicted to have only the weak 
$Q^2$-dependence due to QCD evolution. As $Q^2 \rightarrow 0$, $I_{GDH}(Q^2)$ must change sign in order to reach the negative value predicted by the GDH sum rule at $Q^2 =0$.

Several phenomenological models have been proposed to describe 
the dependence of the generalised GDH
integral on $Q^2$~\cite{ans,teryaev,burkert,azna,ede}. 
Some of these models predict large effects from
nuc\-leon-res\-on\-ance excitation~\cite{burkert,azna} 
or from higher twist~\cite{ans, ede}, even for $Q^2$ up to a few GeV$^2$.
Other models based on chiral perturbation theory have been proposed
but their application is limited to $Q^2 \ll 1$ GeV$^2$~\cite{chi}.

The contribution of the region $W^2\ge 3.24$~GeV$^2$ to the GDH integral 
defined in Eq.~\ref{for2} was recently measured~\cite{gdhdis} 
for the proton and the neutron in the range 0.8~GeV$^2 \le Q^2 \le 12$~GeV$^2$, 
showing that higher-twist effects do not appear to be  
significant in the measured region. 
This paper presents a measurement of the contribution of the resonance region
to the GDH integral for the proton in a similar $Q^2$-range
(1.2~GeV$^2 \le Q^2 \le 12$~GeV$^2$). 
In combination with the analysis at higher $W^2$, this
provides the first experimental determination of essentially the complete 
GDH integral for the proton over a range of $Q^2$ values.

The measurement was performed in 1997 with a 27.56 GeV beam of longitudinally 
polarised positrons incident on a longitudinally polarised $^1$H gas 
target internal to the HERA storage ring at DESY.
The beam polarisation was measured continuously using Compton
backscattering of circularly polarised laser light~\cite{bar}.
The average beam polarisation for the analysed data was $0.55$.

The HERMES polarised target~\cite{ste} consists of polarised atomic $^1$H
gas confined in a storage cell, which is a
40~cm long open-ended thin-walled 
elliptical tube located on the beam axis inside the HERA vacuum pipe. 
It is fed by an atomic-beam source of nuclear-polarised hydrogen based
on Stern-Gerlach separation~\cite{sto}.
It provides an areal target density of about  
$7 \times 10^{13}$ atoms/cm$^2$.
The nuclear polarisation of atoms and the atomic fraction are
continuously measured with a Breit-Rabi polarimeter and a target
gas analyser~\cite{BRP}. 
The average value of the target polarisation for the analysed data was
$0.88$~\cite{g1p}. The fractional systematic uncertainties of the beam and target
polarisations were 3.4\% and 4.5\%, respectively.
The integrated luminosity for this data set was 70 pb$^{-1}$. 

Scattered positrons were detected by the HERMES 
forward spectrometer, which is described in detail elsewhere~\cite{hspect}. 
The kinematic requirements on the scattered positrons for
the analysis in the nuc\-leon-res\-on\-ance region were:
1~GeV$^2 < W^2 < 4.2$~GeV$^2$, 1.2~GeV$^2 < Q^2 < 12$~GeV$^2$. 
After applying data 
quality criteria, about 0.13~million events were selected.

\begin{figure}[t!]
\begin{center}
\epsfig{file=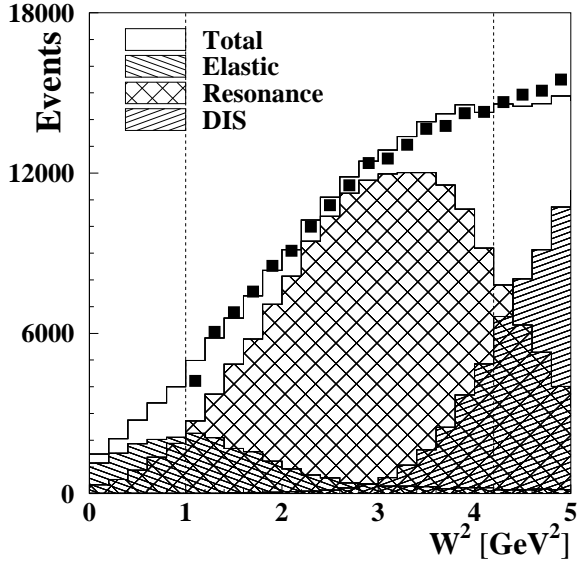,width=8.0cm}  
\end{center}
\caption{Comparison of the event distribution for $W^2 > 1$~GeV$^2$ 
(squares) with the Monte-Carlo simulation (histogram). An overall normalisation factor was applied to the simulation to match the data.
Also shown are the smeared distributions from the elastic, resonance and the deep
inelastic regions. The vertical lines indicate
the resonance region considered in the analysis.} 
\label{yield}
\end{figure}

For all detected positrons the angular resolution was better than 0.6~mrad, the
momentum resolution was better than 1.6\% aside from bremsstrahlung
tails and the $Q^2$-resolution was better than 2.2\%.  
The limited $W^2$-resolution (about 840 MeV$^2$, or $\Delta W \simeq$ 
240~MeV) in the resonance region did not allow 
the  contributions of the individual nucleon resonances to be distinguished.
To evaluate the smearing corrections and the contaminations intruding into the resonance region from the elastic and deep-inelastic
regions, events were simulated using a Monte-Carlo code 
that includes elastic, deep-inelastic and resonance contributions.
The description of the resonance contribution was based on the model 
of Ref.~\cite{bodek}. The deep-inelastic region was modelled using the 
parameterisation of Ref.~\cite{NMCP8} while the elastic form factors were taken
 from Ref.~\cite{FF}. 
In Fig.~\ref{yield} the distribution of events as a function of $W^2$
is presented in comparison with the simulation. It is apparent that the shape 
of the simulated distribution agrees well with that of the data.
It was found that the relative contaminations from the elastic and DIS regions 
in the yield of the resonance region range from 10\% to 2\% and from 7\% to 
16\% respectively, as $Q^2$ increases from 1.2~GeV$^2$ up to 12~GeV$^2$.

Data were divided into six bins in $Q^2$, but only one bin
in $W^2$.
In each $Q^2$-bin the average
longitudinal asymmetry $A_1$ for virtual photoabsorption was calculated using
the formula
\begin{equation}
A_1=\frac{A_\|}{D} - \eta A_2,
\label{a1}
\end{equation}
where $D$ and $\eta$ are factors~\cite{gdhdis}  
that depend on kinematic variables.
The quantity $D$ depends also on  $R= \sigma_L/\sigma_T$, 
the ratio of the absorption cross sections for longitudinal and transverse
virtual photons. 
$A_2$ is related to longitudinal-transverse interference.
The cross section asymmetry $A_{\|}$ is given by : 
\begin{equation}
A_{\|}=\frac{N^\lant L^\lpar - N^\lpar L^\lant }{N^\lant L^\lpar_P + N^\lpar L^\lant _P}.
\label{ap}
\end{equation} 
Here, $N^\lpar $($N^\lant $) is the number of scattered positrons
for target spin parallel (anti-parallel) to the beam spin
orientation. The deadtime-corrected luminosities for each target
spin state are $L^{\lpar(\lant)}$  and $L_P^{\lpar(\lant)}$, the latter being  
weighted by the product of beam and target polarisations. 
The cross section asymmetry $A_{\|}^{res}$ in the resonance region was corrected for
contaminations originating from elastic and deep-inelastic scattering,
as discussed above.
The asymmetry for the elastic contribution was taken from~\cite{FF}, while for 
the DIS region a parameterisation~\cite{naga} based on world data has been used.
 Model dependent uncertainties due to these asymmetries 
and contributions from the Monte Carlo simulation are negligible, 
as this correction is less
than 5\% for $Q^2<$ 5 GeV$^2$.
Radiative corrections were calculated using the codes described in 
Ref.~\cite{polrad}, and were found not to exceed 2\% of the  
asymmetry $A_{\|}^{res}$.
The values for $A_1^{res}+\eta A_2^{res}$ for the measured range of $Q^2$ 
values are presented in Table 1.
The asymmetry $A_1$ was evaluated using Eq.~\ref{a1} under the 
assumption that $A_2=0.06$~\cite{SLAC} in the whole resonance region, and with an average depolarisation
factor $D$ weighted by the event distribution.

The contribution $I_{GDH}^{res}$ of the resonance region to the GDH integral
was determined in each $Q^2$-bin from the asymmetry $A_1$, 
according to 
Eq.~\ref{for2b}, 
in which the integration limits were determined
by the 1~GeV$^2<W^2<4.2$~GeV$^2$ range.
The unpolarised structure function 
$F_1 = F_2(1+\gamma^2)/(2x(1+R))$ was calculated from a modification
of the parameterisation of $F_2$ given in Ref.~\cite{bodek} 
that accounts for nuc\-leon-res\-on\-ance excitation
assuming $R = \sigma_L/\sigma_T$ is constant and equal to 0.18
in the whole resonance region. 
It is worth noting that due to cancellation between the R--dependences of $F_1$ and $D$ at low $y$, the final result is 
insensitive to the choice of $R$. In the integration the $W^2$ dependence of the integrand $F_1/K$ within the individual bins was fully accounted for.

\begin{figure}[t!]
\begin{center}
\epsfig{file=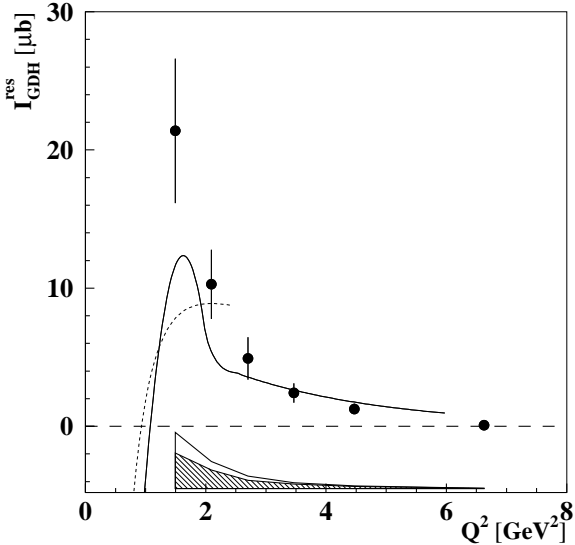,width=8.0cm}
\end{center}
\caption{The GDH integral as a function of $Q^2$ for the region 
1.0~GeV$^2 <W^2 <4.2$~GeV$^2$. The error bars show the statistical 
uncertainties.
The white and the hatched bands represent the systematic uncertainties
with and without the $A_2$ uncertainty contribution. 
The dashed~\protect\cite{burkert} 
and the solid~\protect\cite{azna} curves are predictions based
on a $Q^2$-evolution of nuc\-leon-res\-on\-ance amplitudes.} 
\label{gdh-res}
\end{figure}

The results for $I_{GDH}^{res}$ are presented in Fig.~\ref{gdh-res}. 
The integral strongly decreases with $Q^2$ 
over the entire measured range. 
The magnitude of the systematic uncertainty is indicated by the band.
The dominant contribution to the systematic uncertainty 
is due to the lack of knowledge of $A_2$. 
This contribution  (up to 15\%) was evaluated 
from the total error quoted for a measurement in the resonance region:
$A_2 = 0.06 \pm 0.16$~\cite{SLAC}. This range is consistent with  two other possible assumptions for $A_2$: 
 $A_2=0$, or $A_2 = 0.53 M x /\!\sqrt{Q^2}$, which describes the behaviour in
the deep-inelastic region~\cite{g1p}.
Other contributions are uncertainties from the beam and target polarisations (5.3\%), 
from the spectrometer geometry (2.5\%),
from the combined smearing and radiative  effects (up to 10\%) and 
from the knowledge of $F_2$ (2\%).
The smearing  contribution to the systematic uncertainty 
was evaluated by comparing simulated results from two very different 
assumptions for $A_1$: a power law ($A_1 = x^{0.727}$)
that smoothly extends the DIS behaviour for the asymmetry 
into the resonance region~\cite{naga}, and a step function 
($A_1=-0.5$ for $W^2<1.8$~GeV$^2$ and $A_1=+1.0$ for 1.8~GeV$^2 < W^2 < 4.2$~GeV$^2$) 
that is suggested by the hypothesis of the possible dominance of the 
$P_{33}$-resonance at low $W^2$ and of the $S_{11}$-resonance at higher $W^2$.

The results for $I_{GDH}^{res}$  are compared in Fig.~\ref{gdh-res} with 
two predictions for the contribution of 
nuc\-leon-res\-on\-ance excitation to the integral defined in Eq.~\ref{for2}.
Burkert and Li~\cite{burkert} 
parameterised the experimental $Q^2$-evolution of the main nucleon resonances 
($P_{33}$(1232), $P_{11}$(1440), $S_{11}$(1535), 
$D_{13}$(1520), $F_{15}$(1680)), and assumed single-quark transitions 
to evaluate the contributions from other resonances.
Aznauryan~\cite{azna}  
described the resonance excitation in the
approximation of infinitely narrow resonances, and included a
contribution from one-pion
exchange in the near-threshold region.
Both models predict a sudden decline in $I_{GDH}^{res}$ as $Q^2$ falls 
below 1.5~GeV$^2$, due to a large negative contribution at low $Q^2$
by the helicity structure of the $P_{33}$-resonance. 
At higher $Q^2$ the $P_{33}$-resonance magnetic 
form factor strongly decreases with increasing $Q^2$, 
and the positive contribution to $I_{GDH}^{res}$ arising from 
the excitation of higher-mass resonances becomes dominant. 
Neither of these models includes
the non-resonant
multi-hadron production channels, which should provide an additional positive 
contribution for the region  $W^2\le$ 4.2~GeV$^2$.
Comparison with the data suggests that for $Q^2 \simeq$ 1.5~GeV$^2$,
the resonance-excitation models are not
sufficient to fully explain the experimental result for $I_{GDH}^{res}$.  
Other predictions exist for the resonance-excitation contribution to generalised
GDH integrals, but they are limited to regions of lower $Q^2$~\cite{ma}.

To complete the evaluation of the full integral $I_{GDH}$,
data from the DIS region (4.2~GeV$^2 < W^2 < 45$~GeV$^2$) were reanalysed 
in the same $Q^2$-bins as for the kinematically more restricted resonance region, 
following the procedure described in a previous HERMES publication~\cite{gdhdis}.
A total of 1.52~million events were selected in this
$W^2$-range. The systematic uncertainty for this region is the same as
published in~\cite{gdhdis,g1p}. The systematic uncertainty  on $A_2$ in DIS region does not contribute significantly. 

In Table 1 the resonance region contribution $I_{GDH}^{res}$, the integrals 
$I_{GDH}^{meas}$ in the full measured region and the full
GDH integrals are reported.  
The latter was calculated in each $Q^2$-interval by adding to $I_{GDH}^{meas}$ 
an estimate of the unmeasured contribution for $W^2 > 45$~GeV$^2$. 
This was calculated using
a multiple-Reggeon exchange parameterisation~\cite{nikolo} for 
$\sigma_{\frac{1}{2}}( \nu, Q^2 )-\sigma_{\frac{3}{2}}( \nu, Q^2 )$ at high energy,
and amounted to about $3.5\,\mu$b for all $Q^2$-bins. 
A parameterisation for $g_1$~\cite{grvs}
based on a NLO-QCD analysis provided within
5\% the same results as the multiple-Reggeon exchange analysis. This 
difference was taken as the  systematic uncertainty 
of the high energy contribution.
It is worth noting that for the low $Q^2$-bins, both the statistical
and the systematic uncertainties of $I_{GDH}$ are dominated by
those from the resonance region.
In this region, these 
uncertainties are large due to the smallness of $D$ 
and to the large size of $\eta$, respectively. 

\begin{figure}[t!]
\begin{center}
\epsfig{file=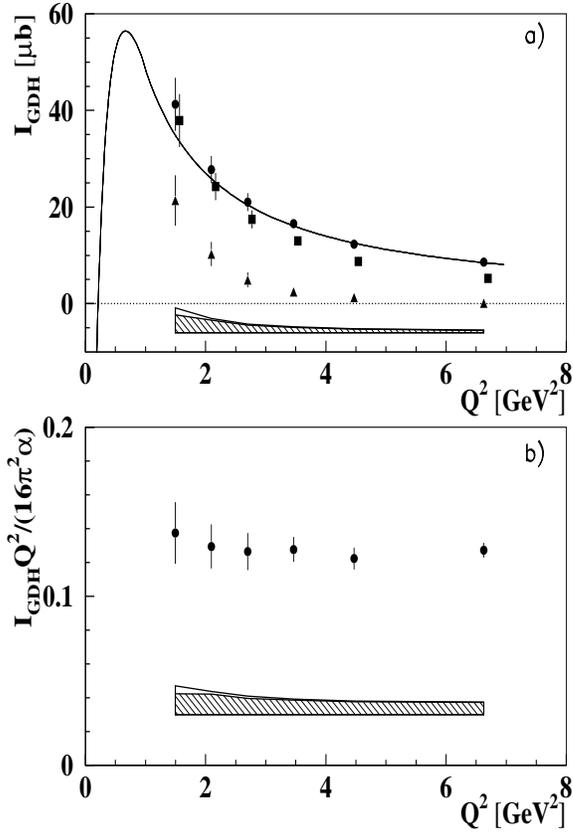, height=12.0cm, width=8.0cm}
\end{center}
\caption{a) $I_{GDH}$ as a function of $Q^2$ for  
various upper limits of integration:
$W^2 \le 4.2$~GeV$^2$ (triangles), $W^2 \le 45$~GeV$^2$ (squares),
and the total integral $I_{GDH}$ (circles). The squares have been slightly
shifted to make them more visible.
The curve is the Soffer-Teryaev model~\protect\cite{teryaev} 
for the total integral. 
b) $I_{GDH} Q^2 /(16 \pi ^2 \alpha)$ as a function of $Q^2$.
For both panels, the error bars show the statistical uncertainties,
and the white and the hatched bands at the bottom represent 
the systematic uncertainties for the total integral with and without 
the $A_2$  contribution.}
\label{gdh-tot}
\end{figure}

In Fig.~\ref{gdh-tot}a) the total GDH integral is shown together
with the partial integrals for $W^2 < 4.2$~GeV$^2$ and for $W^2 < 45$~GeV$^2$.
The contribution of the  resonance region to the full GDH integral is small for $Q^2$ values
above 3~GeV$^2$.

Fig.~\ref{gdh-tot}a) also shows a prediction~\cite{teryaev} 
based on a $Q^2$-evolution of the two polarised structure functions 
$g_1$ and $g_2$, without consideration of any explicit nuc\-leon-res\-on\-ance
contribution. 
This prediction is in good agreement with the experimental data.

In the whole energy range, $I_{GDH}$ is consistent within the uncertainties
($\chi^2/N_{df}=0.4$) with a simple 1/$Q^2$ power law.
This is demonstrated in Fig.~\ref{gdh-tot}b where the results for 
$I_{GDH}$ are multiplied by $Q^2 /(16 \pi^2 \alpha)$. 
In the deep-inelastic limit, this quantity is equivalent to
$\Gamma_1$ (see Eq.~\ref{I1}). 
The present results are in agreement with 
the measurements of $\Gamma_1 = 0.120\pm0.016$ at 
$Q^2=10$~GeV$^2$~\cite{SMC}
and $\Gamma_1 = 0.129\pm0.010$ at $Q^2=5$~GeV$^2$~\cite{SLAC}.
In addition, values of $\Gamma_1$ extracted from the present data are also consistent with a measurement of 
$\Gamma_1 = 0.104\pm0.017$ at $Q^2=1.2$~GeV$^2$~\cite{SLAC} in which the structure
function $g_1$ was  measured in the resonance region.

In summary, the $Q^2$-dependence of the generalised 
Ge\-ra\-si\-mov-Drell-Hearn integral for the proton was determined for the first time 
in both the resonance and the deep-inelastic regions, covering the $Q^2$-range 
from 1.2 to 12~GeV$^2$. 
In the resonance region, the data 
suggest that for $Q^2 \ge$ 1.5~GeV$^2$,
existing resonance-excitation models are not
sufficient to fully explain the experimental result for $I_{GDH}^{res}$.  
Above $Q^2 = 3$~GeV$^2$ the DIS contribution to the generalised
GDH integral is dominant. 
The $Q^2$-behaviour of $I_{GDH}$  suggests that there are no large
effects from either resonances or non-leading-twist, and indicates
that the sign change of $I_{GDH}$ to meet the real photon limit occurs at $Q^2$ lower than 1.2~GeV$^2$.

We thank S. Gerasimov, V. Burkert and I.G. Aznauryan for useful discussions
 and I.G.A. for providing the curves of her calculations.
We gratefully acknowledge the DESY management for its support, the staffs at DESY 
and the collaborating institutions for their significant effort, and our funding
agencies for financial support.

\newpage
\onecolumn
\begin{table}[t!]

Table 1: Results for $A_1^{res}+\eta A_2^{res}$, the resonance part($I_{GDH}^{re
s}$) to
the GDH integral, and the total measured integral
($I_{GDH}^{meas}$), 
as well as the full GDH integral ($I_{GDH}$), including the unmeasured
part .
Errors represent the statistical uncertainty for $A_1^{res}+\eta
A_2^{res}$, and the statistical and the
systematic uncertainties of the integrals. 

\begin{center}
\vskip -0.2cm
\medskip
\begin{tabular}{c@{\hspace{0.1cm}}c@{\hspace{0.1cm}}c@{\hspace{0.1cm}}c@{\hspace
{0.1cm}}c}\hline\hline
$<Q^2>$ & $\;$ $A_1^{res}+\eta A_2^{res}$ &$\;$ $I_{GDH}^{res}$ &$\;$ $I_{GDH}^{
meas}$ &$\;$ $I_{GDH}$ \\
$[$GeV$^2$]&       &$\;$ [$\mu$b]&$\;$ [$\mu$b]&$\;$ [$\mu$b] \\
\hline
  1.5      & 0.71$\pm$0.16  & 21.4$\pm$5.2$\pm$4.1&$\;$37.9$\pm$5.5$\pm$5.1 &$\;
$$\;$41.2$\pm$5.5$\pm$5.1 \\
  2.1      & 0.77$\pm$0.18  & 10.3$\pm$2.5$\pm$2.0&$\;$24.3$\pm$2.8$\pm$2.9 &$\;
$$\;$27.8$\pm$2.8$\pm$2.9 \\
  2.7      & 0.74$\pm$0.22  &$\;$4.9$\pm$1.5$\pm$0.9&$\;$17.5$\pm$1.8$\pm$1.8 &$
\;$$\;$21.0$\pm$1.8$\pm$1.8 \\
  3.5      & 0.79$\pm$0.22  &$\;$2.4$\pm$0.7$\pm$0.4&$\;$13.0$\pm$1.0$\pm$1.2 &$
\;$$\;$16.5$\pm$1.0$\pm$1.2 \\
  4.5      & 0.97$\pm$0.29  &$\;$1.3$\pm$0.4$\pm$0.2&$\;$$\;$8.8$\pm$0.6$\pm$0.8
 &$\;$$\;$12.3$\pm$0.6$\pm$0.8 \\
  6.6      & 0.55$\pm$0.23  &0.08$\pm$0.03$\pm$0.01&$\;$$\;$5.3$\pm$0.3$\pm$0.5 
&$\;$$\;$$\;$8.6$\pm$0.3$\pm$0.5 \\
\hline\hline
\end{tabular}
\end{center}
\end{table}

\end{document}